\documentclass[fleqn,twoside]{article}
\usepackage{espcrc2}

\usepackage{graphicx}

\newcommand{\be}{ \begin{equation}}
\newcommand{\ee}{ \end{equation}  }
\newcommand{\bea}{ \begin{eqnarray}}
\newcommand{\eea}{ \end{eqnarray}  }
\newcommand{\re}{ \mbox{\rm e} }

\title{Dynamical fermions by global acceptance 
       steps\thanks{Presented by U. Wolff}}

\author{Francesco Knechtli\address[HU]{
        Institut f\"ur Physik, Humboldt Universit\"at,
        Newtonstr. 15, 12489 Berlin, Germany} 
        and Ulli Wolff\addressmark[HU]
        }
       
\begin{document}

\begin{abstract}
A study of principle is conducted on the inclusion of the fermionic
determinant as a Metropolis acceptance correction. It is carried out
in the 2-D Schwinger model to prepare
later applications to the Schr\"odinger functional.
A mixed stochastic/determistic acceptance step is found that allows to
include some problematic modes in a way to avoid the collapse of
the acceptance rate due to fluctuations.
\vspace{1pc}
\end{abstract}

\maketitle

\section{Introduction}

The inclusion of light dynamical fermions in a formerly quenched
simulation boosts the computational demand by a very large factor
for standard Hybrid Monte Carlo (HMC) type algorithms.
This is even true in cases like the Schr\"odinger functional at weak
coupling (small physical volume), where we expect the fermionic determinant
to be less influential. We here explore the naive idea to propose in such cases
gauge configurations by an effcient pure gauge update scheme and to filter
these proposals by an accept reject step that leads to the correct
full QCD ensemble. Although it is advantageous to optimize the coupling
or even the type of action to be used for the proposals, we here focus
on the case of proposing with just the gauge part of the full action.
A more detailed account is given in \cite{Knechtli:2003yt}.

\begin{figure}[htb]
\includegraphics[width=17pc]{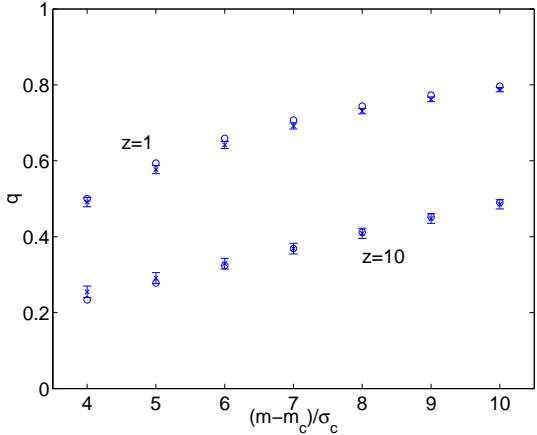}
\caption{Acceptance with the exact determinant, $L=16$.}
\label{qvsm}
\end{figure}

We use the two-flavour two-dimensional Schwinger model 
with a non-compact lattice gauge field as a test-case
with the partition function
\be
Z = \int D \mu(A) \, |\det(D_W + m)|^2 ,
\ee
where $D_W$ is the Wilson Dirac operator.
The gauge action is included in the normalized measure
\be
D\mu(A)
% = \frac1{Z_G} \, 
\propto\prod_{x\mu}dA_\mu(x) 
\, \prod_\mu \delta\bigl(\sum_x A_\mu\bigr)
\; \re^{-S_G - S_\xi} \\
%\; \exp[-S_G - S_\xi]\, ,
\ee
where $S_G$ is the plaquette action and 
\be
S_\xi=\frac1{2\xi^2} \sum_x (\Delta^*_\mu A_\mu)^2
\ee
a gauge fixing term with the backward difference operator $\Delta^*_\mu$. 
Periodic boundary conditions are assumed
and the $\delta$-function eliminates the integration
over two zero-momentum modes. In $D_W$ the compact gauge
variables $U_\mu=\exp(igA_\mu)$ and with them the coupling strength $g$
enter. In the following it is replaced by the dimensionless
variable $z=\sqrt{\sigma}L$, where
$\sigma=g^2/2$ is the string tension.
The gauge part of the action is purely Gaussian.
Gauge fields distributed according to $D\mu(A)$
can be generated in momentum space and then Fourier transformed.
This amounts to an ideal independent sampling or global
heatbath algorithm. In Fig.\ref{qvsm} the unquenched
Metropolis acceptance rate
\bea
 q &=& \frac1Z \int D\mu(A) \int D\mu(A') \\
   & & |\det(D_W+m)|^2  \, w(A,A') \nonumber
\eea
with
\be
w(A,A')=\min\left( 1, \frac{|\det(D'_W+m)|^2}{|\det(D_W+m)|^2} \right)
\ee
is shown for such proposals
as a function of the mass. The determinants have been evaluated
exactly in this case (data with error bars).
The critical mass is defined from the spectral gap 
for individual configurations
and $m_c,\sigma_c^2$ are its mean and variance with respect to
$D\mu(A)$. For the range shown no execptional configurations
are proposed in practice. 
The circles in the plot are obtained by assuming a Gaussian
distribution for the fermionic part of the action,
measuring its width and evaluating $q$ analytically for this model.
Although the cost to compute exact
determinants in higher dimension is prohibitive we here see
that good acceptance of our global correction is possible.
In \cite{Knechtli:2003yt} it is shown that $q$ is actually an upper bound
for the stochastic method discussed next.

\section{Stochastic acceptance steps}

An independent new configuration $A'$ proposed as a successor to $A$
is now accepted with the probability
\be
w_0(A,A';\eta) = \min[1,\rho(M\eta)/\rho(\eta)]
\ee
where $\eta$ is a pseudofermionic random field
generated with some distribution $\rho$
(typically Gaussian) and
$M = (D_W'+m)^{-1}(D_W+m)$ is the `ratio'-operator.
The main cost of this step is just an inversion.
It leads to a correct algorithm due to detailed balance holding
on average,
\bea
 |\det(D_W+m)|^2 \left< w_0(A,A';\eta) \right>_{\eta} &=&
\nonumber\\
 |\det(D_W'+m)|^2 \left< w_0(A',A;\eta) \right>_{\eta}&, &
\eea
which is shown by a change of variables in the $\eta$-integral.
As mentioned before
\be
\langle w_0 \rangle_{A,A',\eta} = q_0 \le q
\ee
has been derived.
It turns out that for most of the physical situations
shown in Fig.\ref{qvsm} the value of
$q_0$ is smaller than $q$
to such an extent that the stochastic algorithm is 
rendered useless. We found that this is related to the
spectrum of $M^\dag M$ whos eigenvalues $\lambda_i$ also occur
in the generalized problem with $(D_W+m)(D_W+m)^{\dagger}$
on one side and its primed companion on the other.
Full spectra are easily computed in our model and 
typically show a structure of two pairs of eigenvalues
of order $g^2, g^{-2}$ respectively and the bulk
of the spectrum near one. The occurence of
these special modes can be understood in perturbation theory.
By analytically computing 
$\left< w_0(A,A';\eta) \right>_{\eta}$ as a function of $\{\lambda_i\}$
one sees that these extremal eigenvalues spoil the stochastic
acceptance, even if $\prod \lambda_i \approx 1$ and $q$
is close to one, and then
lead to  $q_0$ being very much smaller than  $q$.
As a remedy to this problem we found that detailed
balance also holds exactly \cite{Knechtli:2003yt}
for a partially stochastic (PSD) estimation of the determinant ratio
with the acceptance probability
\bea
&& w_4(A,A') = \\
&& \min\left[1,\prod_{i\in S} \lambda_i^{-1}
\exp(-\eta^\dag \bar{P}(M^\dag M-1)\bar{P}\eta)\right].
\nonumber
\eea
Here $S$ is the index set corresponding to the four extremal eigenvalues
$\lambda_i$ (two from either end of the spectrum) 
that caused the trouble. The complement of the span
of the corresponding eigenvectors defines the
projector $\bar{P}$. Instead of four also any other even
number of extremal eigenvalues may be taken
symmetrically from both ends of the spectrum,
and then
this family of algorithms interpolates between
the exact determinant $w$ and the fully stochastic method with $w_0$.
A few extremal generalized
eigenvalues and -vectors may for instance be
computed with Lanczos or Ritz minimization methods, but this
point needs further study.

\begin{figure}[htb]
\includegraphics[width=17pc]{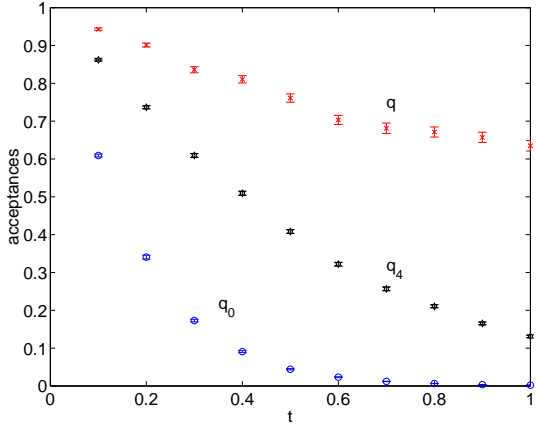}
\caption{Acceptance vs. stepsize $t$ ($z=2,L=8,m=m_c+5\sigma_c,\xi=0$).}
\label{accvst}
\end{figure}

\begin{figure}[htb]
\includegraphics[width=17pc]{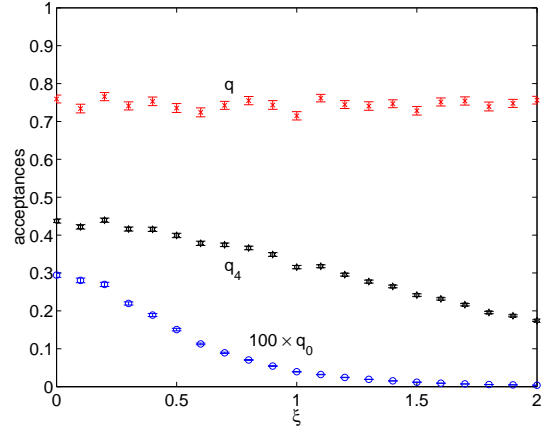}
\caption{Acceptance vs. gaugeparameter $\xi$ ($z=1,L=8,m=m_c+8\sigma_c,t=1$).}
\label{accvsxi}
\end{figure}

Results for acceptance rates are found in Figs.\ref{accvst} and \ref{accvsxi}.
Here a free parameter $t$ has been introduced,
that allows so vary the size of proposed
moves $A \to A''=\cos(\pi t/2) A +\sin(\pi t/2) A'$ 
with $A'$ generated by global heatbath. Also these proposals
obey detailed balance with respect to the gauge action.
Fig.\ref{accvst} demonstrates that for larger $t$-values
PSD remains a feasible algorithm. 
As long as the gauge parameter $\xi$ vanishes, 
simulated fields occur
in a completely fixed gauge, while for growing $\xi$ the proposals
are positioned more and more randomly on their gauge orbits.
Fig.\ref{accvsxi} shows as one may intuitively expect that
this freedom of gauge mismatch
lowers the acceptance rate. Note that $q$ with the
exact determinant is independent of $\xi$. 
For a locally gauge invariant distribution $\rho$
the $\eta$-averages
of $w_0$ and $w_4$ are invariant under transforming both $A$ and $A'$
with the same gauge function due to the gauge covariance of $D_W$
but not under gauge moves of $A'$ relative to $A$. Hence the acceptance
rate and autocorrelation times
are $\xi$-dependent although physical results are not.

\section{Conclusions}

We found that at least in the 2-D Schwinger model the inclusion
of fermions by a global accept step is possible. Gauge fields
in this superrenormalizable model are generally smoother than in QCD
and this is probably important for the success of the method.
In sufficiently small physical volume with Schr\"odinger functional
boundary conditions however a similar situation arises. 
We hope to be able to apply the global acceptance method there
and maybe find an efficiency superior to HMC.
This method has already found 4-D applications with blocked
and hence smoother gauge fields entering into the Dirac operator
\cite{Hasenfratz:2002vv,Hasenfratz:2002jn}. 
Here the use of HMC seems too complicated
due to the complex dependence of the fermion action on the fundamental
(unblocked) gauge fields.
Also for such applications it seems to be valuable
to understand the behaviour of the acceptance rate 
in detail and to
look for tricks to boost it.

\bibliography{lattice}           %or whatever your .bib file is
\bibliographystyle{h-elsevier}   %if you use h-elsevier.bst

\end{document}